\documentclass[useAMS,usenatbib,letters]{mn2e}
\usepackage{graphicx}
\usepackage{aasabbrev}

\newcommand{\unit}[1]{\,\hbox{#1}}

\title[Local Black Hole Scaling Relations Imply Compton Thick or Super
Eddington Accretion]{Local Black Hole Scaling Relations Imply Compton
  Thick or Super Eddington Accretion}

\author[G. S. Novak]{G. S. Novak$^{1}$\thanks{E-mail: greg.novak@obspm.fr}\\
$^{1}$Observatoire de Paris, LERMA, CNRS, 61 Av de l'Observatoire,
75014, Paris, France
}

\begin{document}

\date{Accepted \ldots. Received \ldots; in original
  form 2013 October 11}

\pagerange{\pageref{firstpage}--\pageref{lastpage}} \pubyear{2013}

\maketitle

\label{firstpage}

\begin{abstract}
  A recent analysis of black hole scaling relations, used to estimate
  the local mass density in black holes, has indicated that the
  normalization of the scaling relations should be increased by
  approximately a factor of five.  The local black hole mass density
  is connected to the mean radiative efficiency of accretion through
  the time integral of the quasar volume density. The correspondence
  between this estimate of the radiative efficiency and that expected
  theoretically from thin-disk accretion has long been used as an
  argument that most of the growth in black holes occurs via luminous
  accretion.  The increase of the mass density in black holes pushes
  the mean observed radiative efficiency to values below that expected
  for thin-disk accretion for any value of the black hole spin,
  including retrograde accretion disks.  This can be accommodated via
  black hole growth channels that are intrinsically radiatively
  inefficient, such as super-Eddington accretion, or via growth
  channels that are intrinsically radiatively efficient but for which
  few of the photons are observed, such as Compton thick accretion.
  Measurements of the 30 keV peak in the X-ray background indicate a
  significant population of Compton thick sources which can explain
  some, but not all, of the change in the local black hole mass
  density.  If this result is taken as evidence that super-Eddington
  accretion is common, it greatly reduces the tension associated with
  the growth timescale of observed $z=7$ quasars compared to the age
  of the universe at that redshift.
\end{abstract}


\section{Introduction}

Super-massive black holes at the centers of galaxies have long been
known to be closely related to the properties of the galaxies that
they inhabit \citep{richstone:98-abridged, gebhardt:00-abridged, ferrarese:00,
  tremaine:02-abridged, haering:04, novak:06-blackholes, gueltekin:09-abridged,
  mcconnell:13, kormendy:13}.  These scaling relations
provide the best estimate of the local mass density in black holes
\citep{yu:02, shankar:09}.  The radiative efficiency of the accretion
process onto supermassive black holes can be estimated by connecting
the time integral of the quasar luminosity function and the local
black hole mass function.  For many years, the mean radiative efficiencies estimated
in this way \citep{soltan:82, yu:02, shankar:09} corresponded roughly to
the expected radiative efficiencies for a thin, radiatively efficient
accretion disk \citep{shakura:73, novikov:73}, indicating that most of
the mass that falls into black holes over the lifetime of the universe
does so in this fashion.  This is to be contrasted with the
possibility that the flows are radiatively inefficient because the
accretion rates are a small fraction of the Eddington rate
\citep{narayan:94}, radiatively inefficient because the accretion
rates are greater than the Eddington rate \citep{begelman:79,
  abramowicz:80, paczynsky:80}, radiatively efficient but difficult to
observe because the accretion is heavily obscured, or that the present
mass density in black holes is dominated by the initial mass density
of seeds \citep[see][for a discussion of black hole
seeds]{volonteri:08}.  \citet{paczynski:82} argues that it
was this very correspondence between the estimates of the observed
mean radiative efficiency and that expected efficiency for luminous
radiatively efficient Eddington-limited accretion that led to
diminished interest in radiatively inefficient, super-Eddington
accretion models for many years.

Analyses of the X-ray background (XRB) have raised the
possibility of a significant population of Compton-thick sources.  If
confirmed, these sources contribute to the local black hole mass
density without contributing to the population of observed active
galactic nuclei (AGN).  
A recent comprehensive analysis of black hole mass measurements and
scaling relations \citep{kormendy:13} concluded that the normalization
of the black hole mass---bulge mass relation increased from the
previously accepted value of $M_{\rm BH} = 0.1\% M_{\rm bulge}$ to
$M_{\rm BH} = 0.5 \% M_{\rm bulge}$ with a weak dependence on mass.  
That is to say that the missing black hole mass density implied by
the XRB may have been found in the form of revised measurements of
local black holes.  
In fact, the revision to the local black hole mass density may be so
large that Compton-thick sources are insufficient to explain it.  In
this case, radiatively inefficient, possibly super-Eddington accretion
flows may be required.  Evidence that super-Eddington accretion flows
are common would help to alleviate the difficulty in explaining how
the observed $z=7$ quasars \citep{mortlock:11-abridged} grew to their estimated
masses given the short time allowed.



The \citet{soltan:82} argument can be formulated as a continuity
equation for the mass density of black holes in the universe:

\begin{equation}
  \rho_{\bullet} = \rho_{\rm S} - \rho_{\rm GW} + 
  \int (\dot{\rho}_{UO} + \dot{\rho}_{\rm OB}
+ \dot{\rho}_{\rm CT} + \dot{\rho}_{\rm RI} ) \, dt
\label{eq:mass-conservation}
\end{equation}
where $\rho_{\bullet}$ is the present day mass density in black holes,
$\rho_{\rm S}$ is the mass density in black hole seeds, $\rho_{\rm
  GW}$ is the mass equivalent of the energy radiated as gravitational
waves during black hole mergers, $\dot{\rho}_{UO}$ refers to
unobscured accretion, $\dot{\rho}_{\rm OB}$ to obscured
accretion, $\dot{\rho}_{\rm CT}$ to Compton-thick accretion,
and $\dot{\rho}_{\rm RI}$ to unobscured but radiatively
inefficient accretion.  The distinction between obscured and
Compton-thick accretion is that the former is visible in X-rays while
the latter is difficult to detect even in X-rays.  

The contribution of black hole seeds to the total mass budget is
expected to be small because nearly all of the proposed models produce
seeds well below the mass at which the contribution to the local black
hole mass density peaks, $\sim 10^8 M_\odot$.  Black hole mergers
cause individual black holes to grow but reduce the mass {\em density} in black
holes due to the production of graviational radiation.  The energy in
gravitational waves is expected to be a few percent of the initial
rest mass energy of the two black holes \citep{baker:01, lousto:10},
so this correction is expected to be small.

Equation \ref{eq:mass-conservation} can be rewritten in the form of luminosity densities and
radiative efficiencies:

\begin{equation}
  \rho_{\bullet}c^2 
  = \rho_{\rm S}c^2 - \rho_{\rm GW}c^2 +  
  \sum_i \int 
  \frac{(1-\epsilon_i)}{\epsilon_i f_i} \frac{dL_i}{dV} \, dt 
\label{eq:soltan}
\end{equation}
where $i \in \left\{UO, OB, CT, RI \right\}$, 
$\epsilon$ refers to the intrinsic radiative efficiency and $f$
to the fraction of those photons that we observe.  This is done
to separate the cases of obscured or Compton-thick accretion, which
are intrinsically radiatively efficient but difficult to observe, from
true radiatively inefficient accretion flows.  We may surmise that
$\epsilon_{UO} \sim \epsilon_{OB} \sim \epsilon_{CT} \gg
\epsilon_{RI}$ while $f_{UO}, f_{RI} \sim 1$ and $f_{OB}, f_{CT} \ll
1$.  The Soltan argument is essentially that the only term that
contributes significantly to the left hand side of this equation is
the first one inside the integral, corresponding to unobscured
accretion.

\section{Local Black Hole Mass Density}
\label{sec:bhmd}

The correlation between black hole mass and bulge mass was one of the
relationships between black holes and galaxy properties to be studied,
and value of the proportionality constant 0.001 endured until recently
\citep{kormendy:95, mclure:02, marconi:03, haering:04}. 
These black hole mass relations are converted into a local black hole
mass density by assuming that every galaxy hosts a black hole
well-described by the aforementioned relations, and that there are not
many massive black holes outside the centers of galaxies, ejected for
example by recoil from gravitational waves \citep{redmount:89,
  volonteri:07} or by three-body interactions following galaxy mergers
\citep{hut:92, Xu:94, volonteri:03}.

It is very important to realize that the {\em scatter} in the black
hole scaling relations has a large effect on the black hole mass
density.  Because the galaxy mass function falls steeply at the high
mass end, the population of $10^9 M_\odot$ black holes (for example)
contains more objects with black holes that are {\em over}-massive
than {\em under}-massive.  This means that the typical $10^9 M_\odot$
black hole lives in a galaxy smaller than one would expect from simply
applying the mean relation without scatter, hence there are more $10^9
M_\odot$ black holes than one would expect from the mean relations,
therefore increasing the scatter tends to increase the mass density of
black holes.  This effect can be significant, and careful studies of
the black hole population such as \citet{shankar:09} and
\citet{merloni:08} take it into account.

Recently the proportionality constant between black hole mass and
bulge mass has risen by a factor of
approximately five owing to systematic effects in the measurement of
black hole masses, in particular the inclusion of more realistic dark
matter halos in orbit-based models \citep{kormendy:13}.
As long as the scatter in the scaling relation remains the same, as it
has in the updated \citet{kormendy:13} relation, effect of the change
in the normalization of the black hole mass scaling relation by a
factor of five is to increase the local mass density in black holes by
the same factor of five.  

A number of other correlations between black hole mass and
various galaxy properties have been proposed, with varying degrees of
success \citep[e.g.][]{graham:01, mclure:02, marconi:03, bettoni:03},
and one may wonder which of these should be used in estimating the
black hole mass density.  The choice of which correlation is used to
estimate the black hole mass density makes a difference
\citep{shankar:09} and thus follow-up work is necessary to assess the
sensitivity of the local black hole mass density to these changes in
the normalization of the black hole scaling relations.

\section{AGN Luminosity Function}
\label{sec:agnlf}

In order to characterize the history of massive black hole accretion
in the universe, it is necessary to include {\em all} sources,
regardless of obscuration or the waveband in which the source is
observed.  Therefore the function that we wish to construct is most
properly named the active galactic nucleus luminosity function
(AGNLF).  For this purpose, AGN come in approximately three flavors
depending on the difficulty of observing them: 1) Unobscured in the
optical, with column density of hydrogen below approximately $N_H <
10^{21} \unit{cm}^{-2}$, 2) Obscured in the optical but visible in
X-rays, with $10^{21} \unit{cm}^{-2} < N_H < 10^{24} \unit{cm}^{-2}$,
and 3) Compton thick, where the gas surrounding the AGN is optically
thick to Compton scattering, $N_H > 10^{24} \unit{cm}^{-2}$.  The last
class of objects is difficult to observe in any band, except perhaps
the reprocessed thermal infrared emission, which suffers from
degeneracies with non-AGN sources such as extreme starbursts.

\citet{soltan:82} used the observed quasar luminosity function to
infer the mass density in ``dead quasars,'' ie, black holes, but he
lacked an independent estimate of the black hole mass density with
which to compare.  Black hole scaling relations have provided the
latter tool and \citet{yu:02} arrived at $\rho_\bullet = 2.5 \pm 0.4
\times 10^5 (h/0.65)^2 M_\odot \unit{Mpc}^{-3}$.  This gives mean radiative
efficiencies between 0.1 and 0.2 depending on the masses of optically
bright quasars.  \citet{merloni:08} carried out a similar
study allowing for various modes of black hole accretion at a broad
range of Eddington fractions and found smaller radiative efficiencies,
$\bar{\epsilon} = (0.0675 \pm 0.025) (\rho_\bullet / 4.3 \times 10^5
M_\odot \unit{Mpc}^{-3})^{-1}$.  \citet{shankar:09} found
$\rho_\bullet = 3-5.5 \times 10^5 M_\odot \unit{Mpc}^{-3}$ depending
on which scaling relation one uses to compute the mass density.  They
also found low mean radiative efficiencies: $\bar{\epsilon} = 0.065$
using their estimate of the AGNLF, or $\bar{\epsilon} = 0.09$ for the
\citet{hopkins:07} luminosity function.  These authors also allowed
for the possibility of Compton-thick accretion in their model and
found that it was sub-dominant.

\section{X-ray Background}
\label{sec:xrb}

The most direct observational constraint on Compton thick accretion
comes from the XRB.  \citet{fabian:90} pointed out that the observed
peak in the XRB at 30 keV could be naturally explained by a Compton
reflection spectrum, implying a significant population of Compton
thick objects.  Such a large column of obscuring gas suggests extreme
objects and therefore high accretion rates, thus a possibly
significant contribution to the local black hole mass density.

Estimates of the fraction of Compton thick AGN vary wildly.
\citet{risaliti:99} argued that the 30 keV peak requires twice as many
Compton thick sources as obscured sources so that Compton thick AGN
are approximately half of the total population of AGN.
\citet{gilli:07} found that the XRB implies approximately equal
numbers of unobscured, obscured, and Compton thick AGN.  These results
imply that unobserved black hole accretion is from 30\% to a factor of
two larger than observed accretion.
However, \citet{treister:09} were able to fit the XRB with a
negligible Compton thick fraction, and \citet{akylas:12} found that
the Compton thick fraction can be anywhere from 5\% to 50\% depending
on the modeling.
NuSTAR is the first focusing X-ray telescope covering the band from 3
to 79 keV \citep{harrison:13-abridged} and may be able to answer this question
by resolving the 30 keV XRB peak into point sources.  In their first
results, they found ten sources, equally divided between obscured and
unobscured, with no Compton thick sources.  Their 90\% upper limit on
the fraction of Compton thick AGN is 30\%
\citep{alexander:13-abridged}.

While Compton-thick accretion could bring the mean observed radiative
efficiency into line with that expected for thin-disk accretion,
present observational constraints to not seem to permit sufficient
Compton-thick accretion to allow this explanation.

\section{Radiative Efficiency of Accretion}
\label{sec:efficiency}

The previous sections have discussed the observational constraints on
the mean radiative efficiency of accretion.  In this section I review
the theoretical expectations for the radiative efficiency.
Standard theory for radiatively efficient, geometrically thin,
optically thick accretion disks says that radiative efficiency
(defined to be $\epsilon = L / \dot{m} c^2$) is equal to the binding
energy of a particle at the innermost stable circular orbit (ISCO)
\citep{shakura:73, novikov:73}.  This ranges from
$1-\sqrt{25/27}\simeq0.037$ for a maximally rotating black hole with a
retrograde accretion disk to $1-\sqrt{1/3} \simeq 0.423$ for a
maximally rotating black hole with a prograde accretion disk.  
For a non-rotating Schwarzschild black hole, the
value is $1-\sqrt{8/9} \simeq 0.057$.

\begin{figure}
  \centering
  \includegraphics[width=\columnwidth]{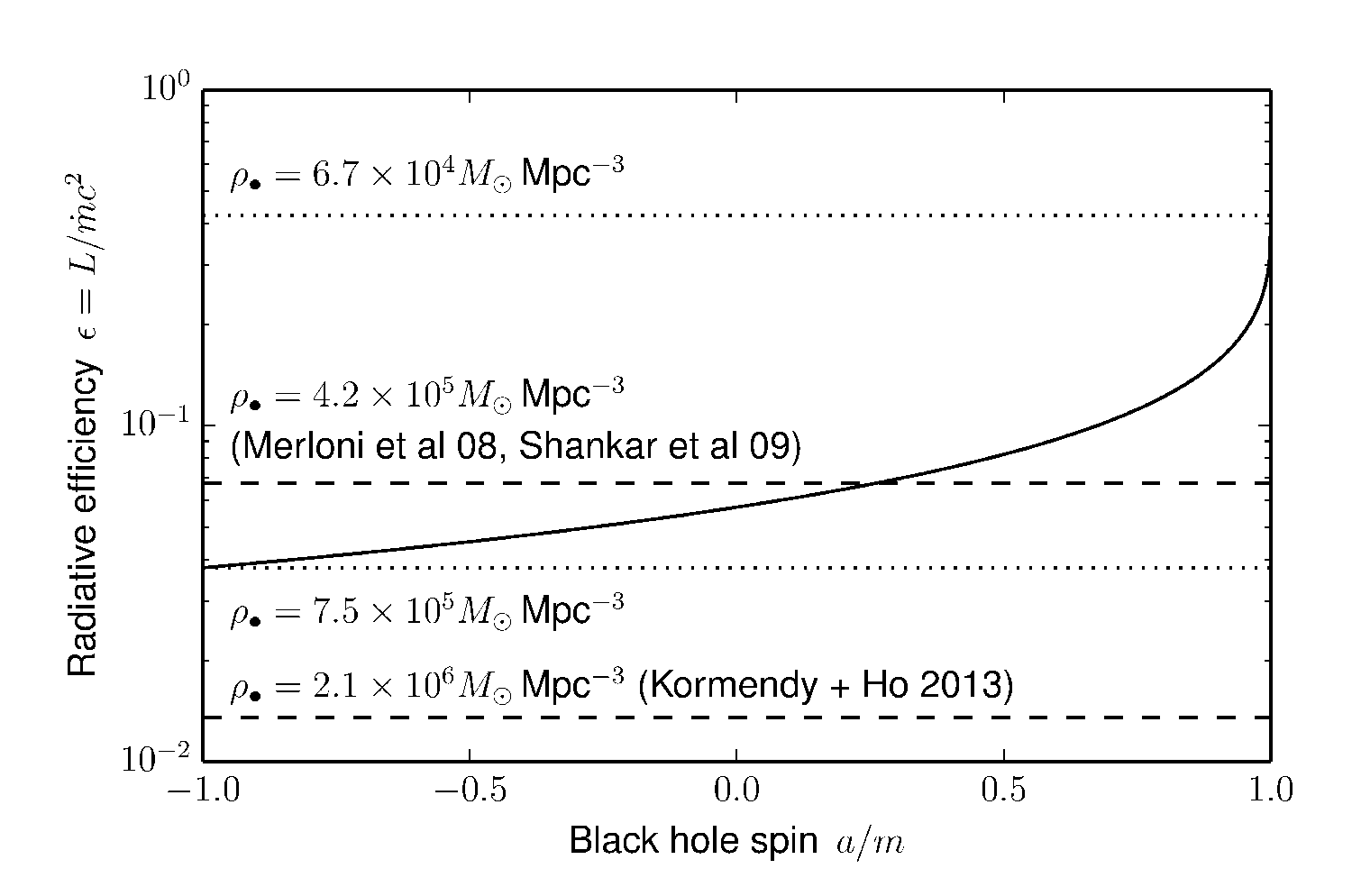}
  \caption{Radiative efficiency as a function of the dimensionless
    black hole spin parameter for radiatively efficient thin-disk
    accretion \citep{shakura:73, novikov:73}.  The spin parameter
    takes values in the range from -1 (maximally spinning black hole
    with a retrograde accretion disk) to 1 (maximally spinning black
    hole with a prograde accretion disk).  Horizontal lines indicate
    the mean radiative efficiencies implied by various values of the
    present day black hole mass density.  Dashed lines indicate
    observational determinations, dotted lines indicate the black hole
    mass density implied by the maximum and minimum theoretical
    efficiencies for thin-disk accretion.  The mean radiative
    efficiency implied by the \citet{kormendy:13} normalization of the
    $M_{\rm BH}$---$M_{\rm bulge}$ relation is below the expected
    efficiency for {\em any} value of the black hole spin, requiring
    significant accretion that is either intrinsically radiatively
    inefficient (e.g. super-Eddington) or radiatively efficient but
    heavily obscured (e.g. Compton-thick).}
  \label{fig-epsilon}
\end{figure}

Figure \ref{fig-epsilon} shows the radiative efficiency as a function
of the dimensionless black hole spin parameter for radiatively
efficient thin-disk accretion along with the mean radiative
efficiencies implied by various values of the present day black hole
mass density.  Negative values of the black hole spin parameter
correspond to retrograde accretion disks.  
The above estimates assume that the stress on the gas is zero in the
so-called plunging region inside the ISCO.  Recent numerical
simulations have indicated that this assumption may not be valid, and
stresses inside the plunging region lead to dissipation that may boost
the radiative efficiency by as much as at 16\% \citep{noble:11}.

The radiative efficiencies found by \citet{merloni:08} and
\citet{shankar:09} using the previously
accepted black hole scaling relations correspond to those
expected for black holes with prograde accretion disks and low to
moderate spins, naturally strengthening the widely held view that most
material that enters black holes does so via a luminous thin-disk
mode.  Given the difficulty in arranging for accreted
material enter the black hole in a predominantly retrograde fashion,
the lower reasonable bound for the mean radiative efficiency is the
value for non-spinning black holes, 5.7\%.  Therefore even a
moderate increase in the local black hole mass density, not to mention
the factor of five increase in the black hole mass density implied by
the \citet{kormendy:13} normalization, pushes the mean radiative
efficiency to values that cannot be reasonably explained in terms of
luminous thin-disk accretion.  This implies that most of the material that
falls into black holes over the history of the universe must do so in
a heavily obscured phase or via a radiatively inefficient
super-Eddington mode.

At accretion rates substantially below the Eddington rate, a variety
of radiatively inefficient accretion models may be relevant
\citep{narayan:94, blandford:99, quataert:00}, but it is difficult to
significantly change the final mass density in black holes precisely
because of the low accretion rates.
Radiatively inefficient accretion is also expected to occur at
accretion rates significantly {\em above} the Eddington rate, owing to
the fact that the flow becomes sufficiently optically thick to advect
photons into the black hole \citep{begelman:79, abramowicz:80,
  paczynsky:80}.  

The other possible explanation for a low observed radiative efficiency
is that it is only the {\em apparent} radiative efficiency that is
low: most of the photons are emitted but do not reach us.  This is the
case for obscured AGN and for Compton-thick AGN.  In both cases, the
intrinsic radiative efficiency is high ($\sim 0.1$), but the fraction
of those photons reaching us is low.

\section{Black Hole Spins}
\label{sec:spins}

Given that the theoretically expected radiative efficiency depends on
the black hole spin and the orientation of the accretion disk, it is
crucial to know whether black holes are spinning rapidly or not.  In
this section I discuss theoretical expectations and observational
constraints on black hole spins.
Angular momentum is expected to be a major barrier to black hole
accretion given the large range in length scales that must be
traversed in order to arrive at the event horizon.  Given the coherent
large-scale angular momentum of many galaxies, one may expect black
holes to be rapidly spinning and to be aligned with the angular
momentum of the gas that the black hole is consuming.

Perfectly coherent gas accretion is not necessary for black holes to
have large spins.  Physical processes such as the \citet{bardeen:75}
effect may be able to align the black hole spin with the angular
momentum of the infalling gas.  For thin-disk accretion the accretion
disk is expected to warp at the radius at a few hundred gravitational
radii where Lense-Thirring precession becomes important, providing a
large lever arm for the infalling gas to exert torques on the black
hole.  While changing the {\em magnitude} of a black hole's spin
requires the black hole to of order double its mass, changing the {\em
  direction} of the spin vector requires the black hole to accrete
only a few percent of its mass \citep{perego:09}.
On the other hand, a picture has emerged recently where accretion is
fully chaotic: gas falls into the black hole in discrete events with
isotropic angular momenta \citep{king:06}.  In this case black hole
spins will be low, with radiative efficiencies of approximately 5\%.

In the case of black hole mergers, the angular momentum of the system
is dominated by the orbital angular momentum for the most interesting
case of approximately equal mass black holes.  The final black hole is
expected to be rapidly spinning \citep{berti:08, lousto:10}, but its
spin direction may not be aligned with the angular momentum of the gas
in the galaxy leading to complicated subsequent evolution.  The
consequences of chaotic versus coherent gas accretion and black hole
mergers in the context of cosmological structure formation have been
studied in detail \citep{dotti:13}.

Measurements of black hole spins will be enormously helpful in
constraining the radiative efficiencies.  While there are a growing
number of spin measurements available, differences in the way
different groups model the X-ray spectra of the same object give
values for the spin that differ at high significance
\citep{brenneman:06, dlCallePerez:10-abridged, patrick:11-one, brenneman:11-abridged,
  patrick:11-two}.

\section{Discussion}
\label{sec:discussion}

Thin-disk accretion predicts mean radiative efficiencies of between
3.7\% (for retrograde accretion disks) and 42\% (for prograde
accretion disks), while 5.7\% is the value associated with
non-spinning black holes.  Given the difficulty in arranging for
accretion to be on average retrograde, the value for non-spinning
black holes is the minimum reasonable value for the mean efficiency if
black hole accretion occurs predominantly via this mode.

If the angular momentum of accreted material is coherent over long
timescales, the black hole spin will eventually become large and
aligned with the angular momentum of the incoming material.  The
requirements for the coherence time are relaxed somewhat if physical
processes such as the \citet{bardeen:75} effect exert sufficient
torque on the black hole to align its spin with the incoming material.
In either case, black hole spins and radiative efficiencies are
expected to be large.  Conversely, chaotic accretion models predict
that black holes will have low spin values, leading to mean radiative
efficiencies around 5.7\%.

Comparison of the local black hole mass function and the AGN
luminosity density over the history of the universe permits a
measurement of the mean radiative efficiency of accretion.  Careful
studies using the previously accepted black hole scaling relation of
$M_{\rm BH} = 0.1\% M_{\rm bulge}$ arrived at mean efficiencies of
around 6.5\%, strengthening the \citet{soltan:82} argument that the
majority of material that falls into black holes through the history
of the universe does so via a luminous thin-disk mode.

However, \citet{kormendy:13} have recently argued that the
normalization of the local black hole scaling relations should be
increased by a factor of five to $M_{\rm BH} = 0.5 \% M_{\rm bulge}$.
This increases the local mass density in black holes by a factor of
five and decreases the mean radiative efficiency by the same factor.
The \citet{shankar:09} estimate goes from $\bar{\epsilon} = 0.065$ to
$0.013$.  This revision puts the mean radiative efficiency far below
the minimum reasonable value of approximately 5.7\%, implying that
most of the material that falls into black holes over the history of
the universe does so in a way that does not produce light that reaches
us.

The low mean radiative efficiency might be because the accretion is
intrinsically radiatively efficient but heavily obscured, i.e. Compton
thick.  The 30 keV peak in the XRB has been taken by some researchers
as an indication of significant Compton-thick accretion---nearly
enough to explain the low radiative efficiencies implied by the
\citet{kormendy:13} result.  However, the most recent data seem to
indicate that it will be difficult to accommodate the full factor of
five via Compton thick accretion.

Low mean radiative efficiencies may also result from intrinsically
radiatively efficient accretion flows.  These are often discussed in
the context of {\em low} accretion rates---below 1\% of the Eddington
rate---but it is difficult to affect the mass budget for black holes
via the behavior of accretion flows at low accretion rates.  If the
explanation is intrinsically radiatively inefficient flows, then it
seems more likely to be evidence for {\em super-}Eddington accretion,
where the radiative efficiency is again expected to fall compared to
thin-disk accretion.  Evidence that super-Eddington accretion is
common would help to alleviate tension associated with the fact that
the growth time for the black holes powering the observed $z=7$
quasars is approaching the age of the universe at that redshift if one
assumes they grow by radiatively efficient accretion.


\section*{Acknowledgments}

G.S.N. was supported by the ERC European Research Council under the
Advanced Grant Program Num 267399-Momentum, and by the National
Science Foundation under Grant No. PHY11-25915 
via the Kavli Institute for Theoretical Physics.  This paper is 
preprint number NSF-KITP-13-234.  The author thanks
Tal Alexander, Arif Babul, Fran\c coise Combes, Fabrice Durier, Kayhan
G\"ultekin, Matthew Lehnert, Francesco Shankar, and Marta Volonteri
for helpful discussions.

\bibliographystyle{mn2e}
\bibliography{common,preprint}

\begin{thebibliography}{}

\bibitem[\protect\citeauthoryear{{Abramowicz}, {Calvani} \&
  {Nobili}}{{Abramowicz} et~al.}{1980}]{abramowicz:80}
{Abramowicz} M.~A.,  {Calvani} M.,    {Nobili} L.,  1980, \apj, 242, 772

\bibitem[\protect\citeauthoryear{{Akylas}, {Georgakakis}, {Georgantopoulos},
  {Brightman} \& {Nandra}}{{Akylas} et~al.}{2012}]{akylas:12}
{Akylas} A.,  {Georgakakis} A.,  {Georgantopoulos} I.,  {Brightman} M.,
  {Nandra} K.,  2012, \aap, 546, A98

\bibitem[\protect\citeauthoryear{{Alexander} et~al.,}{{Alexander}
  et~al.}{2013}]{alexander:13-abridged}
{Alexander} D.~M.,  et~al., 2013, \apj, 773, 125

\bibitem[\protect\citeauthoryear{{Baker}, {Br{\"u}gmann}, {Campanelli},
  {Lousto} \& {Takahashi}}{{Baker} et~al.}{2001}]{baker:01}
{Baker} J.,  {Br{\"u}gmann} B.,  {Campanelli} M.,  {Lousto} C.~O.,
  {Takahashi} R.,  2001, Physical Review Letters, 87, 121103

\bibitem[\protect\citeauthoryear{{Bardeen} \& {Petterson}}{{Bardeen} \&
  {Petterson}}{1975}]{bardeen:75}
{Bardeen} J.~M.,  {Petterson} J.~A.,  1975, \apjl, 195, L65

\bibitem[\protect\citeauthoryear{{Begelman}}{{Begelman}}{1979}]{begelman:79}
{Begelman} M.~C.,  1979, \mnras, 187, 237

\bibitem[\protect\citeauthoryear{{Berti} \& {Volonteri}}{{Berti} \&
  {Volonteri}}{2008}]{berti:08}
{Berti} E.,  {Volonteri} M.,  2008, \apj, 684, 822

\bibitem[\protect\citeauthoryear{{Bettoni}, {Falomo}, {Fasano} \&
  {Govoni}}{{Bettoni} et~al.}{2003}]{bettoni:03}
{Bettoni} D.,  {Falomo} R.,  {Fasano} G.,    {Govoni} F.,  2003, \aap, 399, 869

\bibitem[\protect\citeauthoryear{{Blandford} \& {Begelman}}{{Blandford} \&
  {Begelman}}{1999}]{blandford:99}
{Blandford} R.~D.,  {Begelman} M.~C.,  1999, \mnras, 303, L1

\bibitem[\protect\citeauthoryear{{Brenneman} et~al.,}{{Brenneman}
  et~al.}{2011}]{brenneman:11-abridged}
{Brenneman} L.~W.,  et~al., 2011, \apj, 736, 103

\bibitem[\protect\citeauthoryear{{Brenneman} \& {Reynolds}}{{Brenneman} \&
  {Reynolds}}{2006}]{brenneman:06}
{Brenneman} L.~W.,  {Reynolds} C.~S.,  2006, \apj, 652, 1028

\bibitem[\protect\citeauthoryear{{de La Calle P{\'e}rez} et~al.,}{{de La Calle
  P{\'e}rez}  et~al.}{2010}]{dlCallePerez:10-abridged}
{de La Calle P{\'e}rez} I.,  et~al., 2010, \aap, 524, A50

\bibitem[\protect\citeauthoryear{{Dotti}, {Colpi}, {Pallini}, {Perego} \&
  {Volonteri}}{{Dotti} et~al.}{2013}]{dotti:13}
{Dotti} M.,  {Colpi} M.,  {Pallini} S.,  {Perego} A.,    {Volonteri} M.,  2013,
  \apj, 762, 68

\bibitem[\protect\citeauthoryear{{Fabian}, {George}, {Miyoshi} \&
  {Rees}}{{Fabian} et~al.}{1990}]{fabian:90}
{Fabian} A.~C.,  {George} I.~M.,  {Miyoshi} S.,    {Rees} M.~J.,  1990, \mnras,
  242, 14P

\bibitem[\protect\citeauthoryear{{Ferrarese} \& {Merritt}}{{Ferrarese} \&
  {Merritt}}{2000}]{ferrarese:00}
{Ferrarese} L.,  {Merritt} D.,  2000, \apjl, 539, L9

\bibitem[\protect\citeauthoryear{{Gebhardt} et~al.,}{{Gebhardt}
  et~al.}{2000}]{gebhardt:00-abridged}
{Gebhardt} K.,  et~al., 2000, \apjl, 539, L13

\bibitem[\protect\citeauthoryear{{Gilli}, {Comastri} \& {Hasinger}}{{Gilli}
  et~al.}{2007}]{gilli:07}
{Gilli} R.,  {Comastri} A.,    {Hasinger} G.,  2007, \aap, 463, 79

\bibitem[\protect\citeauthoryear{{Graham}, {Erwin}, {Caon} \&
  {Trujillo}}{{Graham} et~al.}{2001}]{graham:01}
{Graham} A.~W.,  {Erwin} P.,  {Caon} N.,    {Trujillo} I.,  2001, \apjl, 563,
  L11

\bibitem[\protect\citeauthoryear{{G{\"u}ltekin} et~al.,}{{G{\"u}ltekin}
  et~al.}{2009}]{gueltekin:09-abridged}
{G{\"u}ltekin} K.,  et~al., 2009, \apj, 698, 198

\bibitem[\protect\citeauthoryear{{H{\" a}ring} \& {Rix}}{{H{\" a}ring} \&
  {Rix}}{2004}]{haering:04}
{H{\" a}ring} N.,  {Rix} H.,  2004, \apjl, 604, L89

\bibitem[\protect\citeauthoryear{{Harrison} et~al.,}{{Harrison}
  et~al.}{2013}]{harrison:13-abridged}
{Harrison} F.~A.,  et~al., 2013, \apj, 770, 103

\bibitem[\protect\citeauthoryear{{Hopkins}, {Richards} \&
  {Hernquist}}{{Hopkins} et~al.}{2007}]{hopkins:07}
{Hopkins} P.~F.,  {Richards} G.~T.,    {Hernquist} L.,  2007, \apj, 654, 731

\bibitem[\protect\citeauthoryear{{Hut} \& {Rees}}{{Hut} \&
  {Rees}}{1992}]{hut:92}
{Hut} P.,  {Rees} M.~J.,  1992, \mnras, 259, 27P

\bibitem[\protect\citeauthoryear{{King} \& {Pringle}}{{King} \&
  {Pringle}}{2006}]{king:06}
{King} A.~R.,  {Pringle} J.~E.,  2006, \mnras, 373, L90

\bibitem[\protect\citeauthoryear{{Kormendy} \& {Ho}}{{Kormendy} \&
  {Ho}}{2013}]{kormendy:13}
{Kormendy} J.,  {Ho} L.~C.,  2013, \araa, 51, 511

\bibitem[\protect\citeauthoryear{{Kormendy} \& {Richstone}}{{Kormendy} \&
  {Richstone}}{1995}]{kormendy:95}
{Kormendy} J.,  {Richstone} D.,  1995, \araa, 33, 581

\bibitem[\protect\citeauthoryear{{Lousto}, {Nakano}, {Zlochower} \&
  {Campanelli}}{{Lousto} et~al.}{2010}]{lousto:10}
{Lousto} C.~O.,  {Nakano} H.,  {Zlochower} Y.,    {Campanelli} M.,  2010, \prd,
  81, 084023

\bibitem[\protect\citeauthoryear{{Marconi} \& {Hunt}}{{Marconi} \&
  {Hunt}}{2003}]{marconi:03}
{Marconi} A.,  {Hunt} L.~K.,  2003, \apjl, 589, L21

\bibitem[\protect\citeauthoryear{{McConnell} \& {Ma}}{{McConnell} \&
  {Ma}}{2013}]{mcconnell:13}
{McConnell} N.~J.,  {Ma} C.-P.,  2013, \apj, 764, 184

\bibitem[\protect\citeauthoryear{{McLure} \& {Dunlop}}{{McLure} \&
  {Dunlop}}{2002}]{mclure:02}
{McLure} R.~J.,  {Dunlop} J.~S.,  2002, \mnras, 331, 795

\bibitem[\protect\citeauthoryear{{Merloni} \& {Heinz}}{{Merloni} \&
  {Heinz}}{2008}]{merloni:08}
{Merloni} A.,  {Heinz} S.,  2008, \mnras, 388, 1011

\bibitem[\protect\citeauthoryear{{Mortlock} et~al.,}{{Mortlock}
  et~al.}{2011}]{mortlock:11-abridged}
{Mortlock} D.~J.,  et~al., 2011, \nat, 474, 616

\bibitem[\protect\citeauthoryear{{Narayan} \& {Yi}}{{Narayan} \&
  {Yi}}{1994}]{narayan:94}
{Narayan} R.,  {Yi} I.,  1994, \apjl, 428, L13

\bibitem[\protect\citeauthoryear{{Noble}, {Krolik}, {Schnittman} \&
  {Hawley}}{{Noble} et~al.}{2011}]{noble:11}
{Noble} S.~C.,  {Krolik} J.~H.,  {Schnittman} J.~D.,    {Hawley} J.~F.,  2011,
  \apj, 743, 115

\bibitem[\protect\citeauthoryear{{Novak}, {Faber} \& {Dekel}}{{Novak}
  et~al.}{2006}]{novak:06-blackholes}
{Novak} G.~S.,  {Faber} S.~M.,    {Dekel} A.,  2006, \apj, 637, 96

\bibitem[\protect\citeauthoryear{{Novikov} \& {Thorne}}{{Novikov} \&
  {Thorne}}{1973}]{novikov:73}
{Novikov} I.~D.,  {Thorne} K.~S.,  1973, in {Dewitt} C.,  {Dewitt} B.~S.,  eds,
  Black Holes (Les Astres Occlus) {Astrophysics of black holes.}.
pp 343--450

\bibitem[\protect\citeauthoryear{{Paczynski}}{{Paczynski}}{1982}]{paczynski:82}
{Paczynski} B.,  1982, Mitteilungen der Astronomischen Gesellschaft Hamburg,
  57, 27

\bibitem[\protect\citeauthoryear{{Paczy{\'n}sky} \& {Wiita}}{{Paczy{\'n}sky} \&
  {Wiita}}{1980}]{paczynsky:80}
{Paczy{\'n}sky} B.,  {Wiita} P.~J.,  1980, \aap, 88, 23

\bibitem[\protect\citeauthoryear{{Patrick}, {Reeves}, {Lobban}, {Porquet} \&
  {Markowitz}}{{Patrick} et~al.}{2011}]{patrick:11-one}
{Patrick} A.~R.,  {Reeves} J.~N.,  {Lobban} A.~P.,  {Porquet} D.,
  {Markowitz} A.~G.,  2011, \mnras, 416, 2725

\bibitem[\protect\citeauthoryear{{Patrick}, {Reeves}, {Porquet}, {Markowitz},
  {Lobban} \& {Terashima}}{{Patrick} et~al.}{2011}]{patrick:11-two}
{Patrick} A.~R.,  {Reeves} J.~N.,  {Porquet} D.,  {Markowitz} A.~G.,  {Lobban}
  A.~P.,    {Terashima} Y.,  2011, \mnras, 411, 2353

\bibitem[\protect\citeauthoryear{{Perego}, {Dotti}, {Colpi} \&
  {Volonteri}}{{Perego} et~al.}{2009}]{perego:09}
{Perego} A.,  {Dotti} M.,  {Colpi} M.,    {Volonteri} M.,  2009, \mnras, 399,
  2249

\bibitem[\protect\citeauthoryear{{Quataert} \& {Gruzinov}}{{Quataert} \&
  {Gruzinov}}{2000}]{quataert:00}
{Quataert} E.,  {Gruzinov} A.,  2000, \apj, 539, 809

\bibitem[\protect\citeauthoryear{{Redmount} \& {Rees}}{{Redmount} \&
  {Rees}}{1989}]{redmount:89}
{Redmount} I.~H.,  {Rees} M.~J.,  1989, Comments on Astrophysics, 14, 165

\bibitem[\protect\citeauthoryear{{Richstone} et~al.,}{{Richstone}
  et~al.}{1998}]{richstone:98-abridged}
{Richstone} D.,  et~al., 1998, \nat, 395, A14

\bibitem[\protect\citeauthoryear{{Risaliti}, {Maiolino} \&
  {Salvati}}{{Risaliti} et~al.}{1999}]{risaliti:99}
{Risaliti} G.,  {Maiolino} R.,    {Salvati} M.,  1999, \apj, 522, 157

\bibitem[\protect\citeauthoryear{{Shakura} \& {Sunyaev}}{{Shakura} \&
  {Sunyaev}}{1973}]{shakura:73}
{Shakura} N.~I.,  {Sunyaev} R.~A.,  1973, \aap, 24, 337

\bibitem[\protect\citeauthoryear{{Shankar}, {Weinberg} \&
  {Miralda-Escud{\'e}}}{{Shankar} et~al.}{2009}]{shankar:09}
{Shankar} F.,  {Weinberg} D.~H.,    {Miralda-Escud{\'e}} J.,  2009, \apj, 690,
  20

\bibitem[\protect\citeauthoryear{{Soltan}}{{Soltan}}{1982}]{soltan:82}
{Soltan} A.,  1982, \mnras, 200, 115

\bibitem[\protect\citeauthoryear{{Treister}, {Urry} \& {Virani}}{{Treister}
  et~al.}{2009}]{treister:09}
{Treister} E.,  {Urry} C.~M.,    {Virani} S.,  2009, \apj, 696, 110

\bibitem[\protect\citeauthoryear{{Tremaine} et~al.,}{{Tremaine}
  et~al.}{2002}]{tremaine:02-abridged}
{Tremaine} S.,  et~al., 2002, \apj, 574, 740

\bibitem[\protect\citeauthoryear{{Volonteri}}{{Volonteri}}{2007}]{volonteri:07}
{Volonteri} M.,  2007, \apjl, 663, L5

\bibitem[\protect\citeauthoryear{{Volonteri}, {Haardt} \& {Madau}}{{Volonteri}
  et~al.}{2003}]{volonteri:03}
{Volonteri} M.,  {Haardt} F.,    {Madau} P.,  2003, \apj, 582, 559

\bibitem[\protect\citeauthoryear{{Volonteri}, {Lodato} \&
  {Natarajan}}{{Volonteri} et~al.}{2008}]{volonteri:08}
{Volonteri} M.,  {Lodato} G.,    {Natarajan} P.,  2008, \mnras, 383, 1079

\bibitem[\protect\citeauthoryear{{Xu} \& {Ostriker}}{{Xu} \&
  {Ostriker}}{1994}]{Xu:94}
{Xu} G.,  {Ostriker} J.~P.,  1994, \apj, 437, 184

\bibitem[\protect\citeauthoryear{{Yu} \& {Tremaine}}{{Yu} \&
  {Tremaine}}{2002}]{yu:02}
{Yu} Q.,  {Tremaine} S.,  2002, \mnras, 335, 965

\end{thebibliography}

\label{lastpage}

\end{document}